\documentstyle[aps]{revtex}
\begin{document}
\draft
\preprint{  }
\title{In-medium relativistic kinetic theory and nucleon-meson systems}
\author{
Klaus Morawetz}
\address{
Max-Planck-Gesellschaft, Arbeitsgruppe "Theoretische
Vielteilchenphysik"\\
an der Universit\"at Rostock,
18055 Rostock, Germany}
\author{Dietrich Kremp }
\address{
Universit\"at Rostock, Fachbereich Physik, \\Universit\"atsplatz
1,
18055 Rostock 1, Germany}
\date{\today}
\maketitle
\begin{abstract}
Within the  $\sigma-\omega$ model of coupled nucleon-meson systems,
a generalized relativistic Lenard--Balescu--equation is presented
resulting from a relativistic random phase approximation (RRPA).
This provides a systematic derivation of relativistic transport equations
in the frame of nonequilibrium Green's function technique including medium
effects as well as flucuation effects. It contains all possible
processes due to one meson exchange and special attention is kept to
the off--shell character of the particles. As a new feature
of many particle effects, processes
are possible which can be interpreted as particle creation and annihilation due
to in-medium one meson exchange.
In-medium cross sections are obtained from the generalized derivation
of collision integrals, which possess complete crossing
symmetries.
\end{abstract}

\newcommand{\grlo}{\stackrel{>}{<}}
\newcommand{\logr}{\stackrel{<}{>}}

\section{Introduction}\label{ch1}

Since Walecka\protect\cite{SW86} has established his model
of meson exchange already 20 years ago, field theoretical models were
successfully applied
for the description of heavy-ion collisions at intermediate energies
\protect\cite{SG86,BD88,S90,C90}. Thereby the derivation of transport equations
starting from a microscopic model is paid great attention
\protect\cite{BM90,D90,I87}.

The most powerful and appropriate method to describe quantum many
particle systems under extreme conditions like nuclear matter or
condensed stellar objects are the real-time-Green functions
technique. Many excellent reviews are published concerning this
method during the past decade \protect\cite{BM90,D90,D84,IZ80,ksb85}.
Moreover there are many investigations in classical
relativistic treatment \protect\cite{MK79,GLW80}. Several
papers are devoted the formulation
of transport equations in quantum field theory
\protect\cite{BM90,I87,MD90,E87}.

In this paper we like to present briefly a derivation of
relativistic transport equations for the $\sigma-\omega$ model.
The formal development of relativistic transport models is well
known and documented in the literature \protect\cite{Ma93,DP91}.
Here special attention is kept to the description of fluctuations and
the influence of medium effects on the kinetic equation.
Additional reaction channels are opened by the influence of many
particle effects, which would be forbidden in uncorrelated
systems. Further, we derive a decoupling between nucleon and
meson-equations, which result in an effective squared one-meson exchange
potential containing no mixed terms of meson contributions. This
is established by the use of generalized optical theorems.

The outline of the paper is the following. In chapter 2 we review the
fundamental equations of the Yukawa Lagrangian and introduce the
nonequilibrium Green's function technique based on the very
fundamental principle of weakening of initial correlation. The
structure of equations are developed in the Schwinger formalism,
generalized to the set of four nonequilibrium Green's functions as
it was similar done by Bezzerides \protect\cite{BB72} and
recently by Davis \protect\cite{DP91}.
The spectral information and therefore the quasiparticle
properties are discussed in chapter 3 which establish a
quite natural generalization of Bruckner theory by the aspect of
complete particle -- antiparticle symmetry of interacting
quasiparticles \protect\cite{BM90}.

To set up a consistent kinetic theory an equation for the Wigner
distribution function should be available, which includes all
features of relativistic quasiparticle behaviour like
Pauli-blocking, screening and relativistic effects, such as
scattering between particles and antiparticles and pair
creation and destruction processes. The consequent treatment of
many particle effects adopted
in section 4 as it was done in the
nonrelativistic case \protect\cite{KKER86}. The generalization of collision
integrals include the effect of density fluctuations by an
dynamical meson exchange potential and describes completely the
particle-antiparticle symmetry.
Particularly, it contains Pauli blocking
and inelastic processes by the medium. A dynamical
potential enters the equation due to density fluctuations and the equation can
be considered
therefore as
a generalization of quantum mechanical
Boltzmann equation \protect\cite{HSS89} to a Lenard-Balescu-type one. This
is the main result of the present paper and should be
the starting point to kinetic description of hot nuclear matter
instead of the relativistic Vlassow treatment \protect\cite{XYXS90}.
Similar equations have been obtained in T -- Matrix approximation
but without pair creation and dynamically screening by De Boer
\protect\cite{BW5}, De Groot \protect\cite{GLW80}, Botermans
\protect\cite{BM90} and Malfliet \protect\cite{M84}.

{}From the
found collision integrals we derive
in-medium cross--sections, which differ from ordinary Born approximation
by two facts. Firstly, no mixed coupling terms occur between different kinds
of meson contributions. This is due to
the more general approximation of self energies instead of the
decoupling normally used. Secondly, the medium effects the
cross section by an energy shift from the vector-meson exchange
and the renormalized effective mass by the scalar-meson
contribution.

\section{Model and basic equations}\label{ch2}

The Yukawa Lagrangian, which couples scalar and vector meson
fields to fermion fields describes the simplest form of a model
for nuclear matter \protect\cite{SW86}
\begin{eqnarray}\label{eq1}
L=&&-\overline{\Psi}\left(-i\gamma^\mu \partial_{\mu} + \kappa_0
\right)\Psi +
\nonumber\\
&+&\frac{1}{2}\partial_{\mu}\Phi\partial^\mu\Phi -
\frac{1}{2}m_s^2\Phi^2+ \, : \, g_s \overline{\Psi}\Phi\Psi \,:
\nonumber\\
&-& \frac{1}{4}F_{\mu\nu}F^{\mu\nu} +
\frac{1}{2}m^2_v\varphi_\lambda \varphi^\lambda + \,:\,
g_v \overline{\Psi} \gamma_\lambda
\Psi\varphi^\lambda \,: .
\end{eqnarray}
Here $m_s^0, m_v^0$ are the bare scalar and vector meson
mass and $\kappa_0$ is the bare nucleon one respectively.
Further, :: denotes the
normal ordering procedure to ensure the renormalization of baryon
density \protect\cite{Ro69}. After renormalization of the masses
the coupling constants have to be chosen in such a way, that on one hand,
the equilibrium ground state
properties can be fitted, and on the other hand, the
scattering data can be reproduced \protect\cite{S90,KLW87}.
Because of the renormalizability of the model Lagrangian we can
choose for the occuring divergent vacuum terms a standard
procedure resulting in physical masses. This will be shown in the
Appendix for our used selfconsistent Hartree-Fock approximation.
In the following we use therefore already the renormalized fields
and masses.

{}From (\ref{eq1}) the equations of motion read
\begin{equation}\label{eq2}
\left(\Box + m_s^2 \right) \Phi = g_s \,\, : \,\, \overline{\Psi}
\Psi
\,\, : \,\, \equiv j_{\Psi}
\end{equation}
\begin{equation}\label{eq2a}
\left(\Box + m_v^2\right)\varphi_\nu = g_v \,\, : \,\, \overline{\Psi}
\gamma_\nu
\Psi \,\, : \,\,
 \equiv j_{\Psi v}
\end{equation}
\begin{equation}\label{eq2b}
\left(i\gamma \quad \partial_{\mu} - \kappa \right) \Psi =
-g_s \,\, : \, \Psi \Phi \,: \,\, + \,\, g_v
\,\, : \, \gamma^\lambda \varphi_\lambda \Psi \, : \, \equiv j_\varphi.
\end{equation}

With the help of the Green's function $G^s_o, G^v_0$
for the free meson equation (\protect\ref{eq2},
\protect\ref{eq2a}) the mesonic
degree of freedom in eq. (\protect\ref{eq2b}) can be eliminated
\protect\cite{BM90,BPW70,W76}
\begin{equation}\label{pot}
\left(i\gamma \quad \partial_{\mu} - \kappa \right)
\Psi_{\alpha} =
\left [ g_v^2 \; (\gamma_{\mu} )_{\alpha \beta} \; (G^v_0)^{\mu \rho}
\; (\gamma_{\mu} )_{\delta \theta} -g_s^2 G^s_o\right ) {\bar
\Psi}_{\delta} \Psi_{\theta} \Psi_{\beta}
\end{equation}
Here a greek letter indicates the spinor index.
Recalling that $G_o^s=(-k^2+m_s^2)^{-1}$ one may construct from
(\protect\ref{pot}) a 4-vector
potential, which has the structure $U_s-U_v$.
Therefore in any perturbation expression, where the potential
enters squared, one gets mixed coupling terms between the
vector
and scalar mesons. This usually used decoupling overlooks the
problem of {\it meson initial correlations}, which is assumed to be zero
when
the mesonic degrees are eliminated by inverting
their differential equation into an integral
one, which was done using the free Green's function. Otherwise
some initial terms have to occur.

We will proceed in an other way and develop a systematic many
particle theory, which yields expressions without any mixed
coupling terms with respect to the different meson contributions.

The physical properties of the system can be described by means
of Green's functions or correlation functions. For nucleon system we
define
\begin{eqnarray}\label{eq3}
iG &=& \left<T \Psi \overline{\Psi} \right> = iG_{++}  \qquad
iG^< = - \left<\overline{\Psi} \Psi \right> =-iG_{+-}
\nonumber\\
i\overline{G} &=& - \left<T \_ \Psi \overline{\Psi} \right> = iG _{--} \qquad
iG^> = \left< \Psi \overline{\Psi} \right> = iG_{-+}.
\end{eqnarray}
and for the meson fields we introduce the correlation functions
\begin{eqnarray}\label{eq4}
id &=& \left<T\varphi\varphi' \right> - \left<
\varphi \right> \left<\varphi' \right> = id_{++}
\nonumber\\
i \overline{d} &=& - \left<T_{\varphi \varphi'} \right> +
\left<\varphi \right>\left< \varphi'\right> = id_{--} \nonumber\\
id^<  &=& -\left< \varphi' \varphi \right> + \left<\varphi'\right>
\left<\varphi \right> = -id_{+-} \nonumber\\
id^> &=& \left< \varphi \varphi' \right> - \left< \varphi \right>
\left< \varphi' \right> = id_{-+}.
\end{eqnarray}
Here we used the convenient matrix - notation where the $\pm$
terms of the 2 x 2 correlation matrix will be signed by {\it latin}
letters. For the reason of legibility we write down  only the
scalar meson equations in the following. The vector meson Green function carry
spin indices.
All final results will be presented for both, vector and scalar mesons.

Using the definition (\ref{eq3}) and (\ref{eq4})
the equations of motion for the different Green's functions can
be derived.
The equation of motion for the causal one reads e.g.
\begin{equation}\label{eq5}
\left(i\gamma^\mu \, \partial_\mu - \kappa\right) G = \delta(1-1')
+ \frac{1}{i} \left<T j_\varphi \psi \right>
\end{equation}
\begin{equation}\label{eq6}
\left( \Box + m_s \right)d = - \delta \left( 1-1'\right) -
\frac{1}{i}\left<Tj_\psi \varphi'\right>
\end{equation}
with $j_\varphi$ and $j_\Psi$ from eq. (\protect\ref{eq2}).
Up to now we have not specified the
meaning of the average. Solutions, which coincide
to some special averages, have to be chosen by
appropriate boundary conditions.
For equilibrium, which corresponds to the average with the grand
canonical equilibrium density operator, the
famous KMS condition \protect\cite{KB62,KKER86} holds
\begin{equation}\label{eq7}
G^> \left|_{t=0} = \pm e^{\beta \mu}G^< \right|_{t'= -i\beta}.
\end{equation}
In the nonequilibrium situation and for real time Green's functions this
condition is not valid. Especially $G^<$ and $G^>$ are independent
functions. One powerful possibility to derive real time Green
functions is the
condition of weakening of initial correlation \protect\cite{ksb85},
which is to be
expressed in systems with finite densities.
If the condition for the correlation time $\tau_{corr}$ and the mean
free collision time $\tau_{coll}$
\[\tau_{corr} \ll \tau_{coll} \]
is valid, we have a condition for the right sides of (\protect\ref{eq5})
and (\protect\ref{eq6})
\begin{equation}\label{eq8}
\lim_{t\to-\infty} \left<j_\varphi \psi \right> = \left<\psi
\overline{\psi}\right> \left\{ g_s \left<\Phi\right> - g_v
\gamma^\lambda \left<\varphi_\lambda \right> \right\}.
\end{equation}
This is an asymptotic condition, which breaks the time symmetry
and provides irreversible evolution in nonequilibrium systems.

In the next step the correlated self energy $\Sigma_c$ is
introduced formally by subtracting
the mean field parts \protect\ref{eq8}) from the right side of eq.
(\protect\ref{eq5})
\begin{equation}\label{eq9}
\frac{1}{i} \left<Tj_\varphi \psi \right> - \lim_{t\to-\infty}
\left<T j_\varphi \psi \right> \equiv  \int\limits_c \Sigma_{c} G.
\end{equation}
The way of integration $c$ has to be determined in such a way that
(\protect\ref{eq8}) is fulfilled. This can be found by
\begin{eqnarray} \label{way}
\left. \int\limits_c d\bar {1} \Sigma (1,\bar {1}) G(\bar {1},1')
\right. &=&
\left. \int\limits_{-\infty}^{+\infty} d\bar {1} \left \{ \Sigma (1,\bar {1})
G(\bar {1},1') - \Sigma^< (1,\bar {1}) G^> (\bar {1},1') \right
\} \right. \nonumber\\
&=& \sum_{d=\pm} \int \Sigma_{+d} G_{d+}.
\end{eqnarray}
It is easy to see that the boundary condition is fulfilled, since the
contribution
(\protect\ref{way}) vanishes in the limit $t_1'=t_1^{\pm} \rightarrow - \infty
$.
For the case $t_1<t_1'$ (and vice versa) we can write e.g.
\newpage
\begin{eqnarray}
\lefteqn { \left. \int\limits_{-\infty}^{+\infty} d\bar {1} \left \{ \Sigma
(1,\bar {1})
G(\bar {1},1') - \Sigma^< (1,\bar {1}) G^> (\bar {1},1') \right \} \right. =
} \nonumber \\
&=&\int_{-\infty}^{t_1}\Sigma^> (1,\bar {1}) G^< (\bar {1},1')+
\int_{t_1}^{t_1'}\Sigma^< (1,\bar {1}) G^< (\bar {1},1')+ \nonumber \\
&+&\int_{t_1'}^{\infty}\Sigma^< (1,\bar {1}) G^> (\bar {1},1')-
\int_{-\infty}^{\infty}\Sigma^< (1,\bar {1}) G^> (\bar {1},1') \nonumber \\
&& \rightarrow 0 \; for \; t_1 \rightarrow t_1 \rightarrow t_o =-\infty
\end{eqnarray}

If we split the last integral on the right of (\protect\ref{way})
into two parts according to
\[
\int_{-\infty}^{+\infty}d{\bar t}_1=\int_{-\infty}^{t_1'}d{\bar t}_1+
\int_{t_1'}^{+\infty}d{\bar t}_1
\]
a contour of time integration follows which is equal to the
Keldysh--contour \protect\cite{D84,D90}.

Now we are able to enclose the equation of motion
(\protect\ref{eq5}) resulting in the nonequilibrium Dyson equation as
matrix equation (\protect\ref{eq3}) in the following form
\begin{equation}\label{eq11}
\left(i\gamma^\mu \partial_\mu - \kappa \right) G_{bc} (11') =
\delta_{bc} (11') + \int\limits_{-\infty}^{\infty} d \stackrel{=}{1}
\Sigma_{b\overline{a}}\left(1\stackrel{=}{1}
\right)G_{\overline{a}c}\left(\stackrel{=}{1} 1' \right)
\end{equation}
and for the meson correlation functions (\protect\ref{eq6}) one has
\begin{equation}\label{eq12}
\left( \Box + m^2 \right) d_{bc}(1 1') = \delta_{bc} (1 - 1') +
\sum_{d} \int d2 \Pi_{bd} (12) d_{dc}(21').
\end{equation}
Equations (\protect{eq11}) and (\protect\ref{eq12}) are matrix
equations, where latin letters remark (+,-)
in the following. To study the structure of the self energy $\Sigma$ and
the polarization function $\Pi$ explicitly we want to use a technique developed
by
Schwinger \protect\cite{Ro69}. Therefore we introduce an
infinitesimal meson generating flux $j_{\Phi}$ for
scalar mesons and $j_{\lambda}$ for vector mesons, as a new type of
interaction
\begin{equation}\label{eq13}
L_{\mbox{int}} = - j_\Phi \Phi - j_\lambda
\varphi^\lambda.
\end{equation}
In our presented formalism it is not necessary to introduce a
generating functional for nucleons. It turns out that the
infinitesimal interaction (\protect\ref{eq13})
is sufficient to obtain the Kadanoff-Baym \protect\cite{KB62} equations.
Therefore they are valid for any density or correlations in the system.

At this point it has to be
remarked, that we suppress the nondiagonal terms of the vector
mesons by choosing the generating functional in diagonal form.
This is possible, because they must not contribute to physical
observables such as S-matrices as a result of coupling to the baryon conserving
flux. The influence to non -- observable quantities such as
self energy should be considered in principle. But they are
neglected here for the reason of simplicity, which is in
agreement with the treatment of Horowitz and Serot
\protect\cite{HS83}.

Now we introduce an interaction picture in respect to the
infinitesimal interaction (\protect\ref{eq13}).
Further, we distinguish between the upper and lower branch \protect\cite{BB72}
of the contour because in such a way we can find relations for
the time evolution operator of this (infinitesimal)
interaction on the Keldysh contour, which reads
\begin{equation}\label{eq14}
S_c=T_c \exp\left\{-ic \int \left(- j_\Phi \Phi
- j_\mu \varphi ^\mu  \right)_c  \right\}.
\end{equation}
Thus special relations can be established by variational
technique
\begin{equation}\label{eq15}
\frac{\delta S_c}{\delta j_b} =-ib \, T_b \, \varphi' \, S_b \,
\delta_{bc},
\end{equation}
where $j$ stand for $j_{\Phi}$ or $j_{\lambda}$.
Latin letters $b,c$ remark (+,-) respectivly.
In a straight forward manner one can express all correlation functions
(\ref{eq4}) through variations with respect to $j$ \protect\cite{BB72}. The
causal Green's function e.g. can be expressed as
\begin{equation}\label{16}
i\frac{\delta<\varphi(1)>_+}{\delta j_+(1')}=\left<T_+
\varphi(1)\varphi(1')\right>-<\varphi(1)>_+<\varphi(1')>_+
\equiv d_{++}.
\end{equation}
With some manipulations the introduced polarization function
(\protect\ref{eq12}) of mesons
can be derived from (\protect\ref{eq6}) in the form
\begin{equation}\label{eq17}
\Pi_{bd}(12)=-g^2 \, b\sum_{a\overline{a}}\int Tr
\left\{G_{ba}(1\overline{1})\Gamma_{a\overline{a}d}
\left(\overline{1}\stackrel{=}{1}2 \right)G_{\overline{a}b}
\left(\stackrel{=}{1} 1^+ \right) \right\}
\end{equation}
where we have introduced the vertex function
\begin{eqnarray}\label{eq18}
\Gamma_{a\overline{a}d}\left(\bar{1}\stackrel{=}{1} 2  \right)
&=& - \frac{1}{g} \,\, \frac{\delta G_{a\overline{a}}^{-1}
\left(\overline{1} \stackrel{=}{1} \right)}{\delta<\varphi(2)>_d}
\nonumber\\
&=& -i \delta_{a\bar{a}}\left(\overline{1}-\stackrel{=}{1}
\right) \delta_{ad} \left(\overline{1}-2 \right)
+ \frac{1}{g} \,\,
\frac{\delta \sum_{a\overline{a}}\left(\overline{1}
\stackrel{=}{1}\right)}{\delta<\varphi(2)>_d}.
\end{eqnarray}
The second line of (\protect\ref{eq18}) is quite easy to verify by means of
(\ref{eq11}).
Concerning the nucleons we find that the
right hand side of equation (\ref{eq5}) could be expressed by the
help of variations in such a way that the equation
(\ref{eq5}) takes the form
\begin{eqnarray}\label{eq19}
\left[ i\gamma ^\mu\delta_\mu - \kappa_0 + g_s<\phi>_b
-g_v\left<\varphi_{\lambda}\right> \gamma^\lambda \right] G_{bc}(11') =
\delta_{bc}(11') \nonumber\\
+ \left\{\frac{bg_s}{i} \frac{\delta}{\delta j_{\phi b}(1)} -
\frac{bg_v}{i} \gamma_\lambda
\frac{\delta}{\delta j_{\lambda b }(1)} \right\} G_{bc}(11').
\end{eqnarray}
After the introduction of the same vertex function
(\ref{eq18}), the structure of the
self energy introduced in (\protect\ref{eq11}) can be written finally as
\begin{equation}\label{eq20}
\Sigma_{b\bar{a}}(1\stackrel{=}{1}) = -\sum \, \int : \quad
bg^2 G_{ba} (1\bar{1})\Gamma_{a\bar{a}d}
(\overline{1} \stackrel{=}{1} 2)d_{db}(2 \, 1) \quad :.
\end{equation}
Equations (\ref{eq11}), (\ref{eq12}), (\ref{eq17}),
(\ref{eq18}) and (\ref{eq20}) form a complete quantum statistical description
of
the many particle system in nonequilibrium and serves as a starting point for
any
further approximated treatment. We have to point out, that the
different contributions of mesons are linearly additive in the self
energies and therefore quadratic additive in the coupling
constants as it is to be seen from (\ref{eq19}). This differs
from other
treatments, where on the stage of eq. (2)
the different  parts
are decoupled approximately yielding a linear additive coupling. Since we have
not
used any restriction or approximation in deriving the general
structure of (\ref{eq19}), it has to be considered as a more
general treatment. The physical reason is, that during the
derivation of the normal simple additive coupling initial
correlations of different meson are neglected. By the sceme
proposed here this is done carefully resulting in a quadratic
additive behavior.

\section{Spectral information}\label{ch3}

Because a complete description of a nonequilibrium situation
demands {\it two}
independent correlation functions instead of one, as it is the
case in equilibrium,
we have some relations between the {\it four} Green functions
\begin{eqnarray}\label{22}
G^r - G^> = G^a - G^< &=& \overline{G} \nonumber\\
G^r + G^< = G^a + G^> &=& G,
\end{eqnarray}
where we introduce the retarded and advanced functions
\protect\cite{D84}
\begin{equation}\label{eq23}
G^{r/a}= G_\delta \pm \Theta\left(\pm\left(x_0
-x'_0\right)\right) \left\{G^>-G^< \right\}.
\end{equation}
This function determines the spectral information
\protect\cite{KKER86}
and we get the equation for them from
(\ref{eq11})
\begin{equation}\label{eq24}
\left(i\gamma^\mu \partial_{\mu} - \kappa \right) G^r = \delta (1 \,\,
1' ) + \int d {\bar 1} \Sigma^r(1 {\bar 1}) G^r({\bar 1} 1').
\end{equation}
After the Wigner transformation, the variables are splitted
in macroscopic and microscopic ones and it is assumed \protect\cite{Ro69}
\begin{equation}\label{25}
\Sigma^r (X,p) \gg \left| \frac{\partial}{\partial x^\mu}
\frac{\partial}{\partial p_\mu} \Sigma^r(X,p) \right| \gg ... .
\end{equation}
This means that $\Delta x^\mu \Delta p_\mu \gg 1$ where the
characteristic length $\Delta p$, at which the self energy varies
in four--momentum space, corresponds to the inverse space--time
interaction range. Therefore the approximation (\ref{eq26})
demands the shortness of the space--time interaction range when
compared with the systems space--time inhomogeneity scale
\protect\cite{MD90}.

All this is the physical background, when we applied the so
called gradient expansion at this place, which yields now for
(\protect\ref{eq24})
\begin{equation}\label{eq26}
\left( \gamma^\mu p_{\mu} - \kappa - \Sigma^r (pX) \right)
G^r(pX) = 1.
\end{equation}
To make further progress, we write the self energy in characteristic parts in
terms of the
$\gamma-$matrices, which can be verified by
general considerations \protect\cite{FW71,SRS90}
\begin{equation}\label{eq27}
\Sigma^r = \gamma^\mu \Sigma_\mu^r + I \Sigma_I^r.
\end{equation}
Further it is useful to split up the self energy  in
real $Re \Sigma$ and imaginary $\Gamma$ parts,
between which exist the dispersion relation
\begin{equation}\label{eq28}
Re \Sigma^r (pR \omega T)= P \int \frac{d\overline {\omega}}{2\pi}
\frac{\Gamma(p{\bar \omega}RT)}{\omega -
\bar{\omega}}.
\end{equation}
Introducing the following medium dressed variables
\begin{eqnarray}\label{eq29}
\tilde{{\bf P}}_\mu &=& p_\mu - Re \Sigma_\mu \nonumber\\
\tilde{{\bf M}} &=& \kappa + Re \Sigma_I \nonumber\\
\tilde{{\bf G}} &=& \left(\gamma_0 \, p_0 \, \varepsilon +
\frac{\Gamma}{2} \right) \left(\gamma \tilde{{\bf P}} +
\tilde{{\bf M}} \right)
\end{eqnarray}
one obtains the complete spectral function from (\protect\ref{eq26}) as
\begin{equation}\label{eq30}
A = i\left\{G^r(\omega + i\varepsilon ) - G^a(\omega -
i\varepsilon )\right\} = \left(\gamma^{\mu} \tilde{{\bf P}}_\mu + {\bf
I\tilde{M}} \right) \frac{2 \tilde{{\bf G}}}{\left(\tilde{{\bf P}}^2 -
\tilde{{\bf M}}^2 \right)^2 + \tilde{{\bf G}}^2}.
\end{equation}
This derivation underlines the correct expression using
$\epsilon \gamma^0$ instead of $\epsilon$
in the denominator as it was pointed out by P. Henning \protect\cite{PH92}.
In the case of vanishing damping $\Gamma$ we get the spectral
function, which determines the quasiparticle energies in the
quasiparticle picture
\begin{equation}\label{eq31}
A = 2 \pi \left\{ \gamma^\mu \tilde{{\bf P}}_\mu + I \;{\bf \tilde{M}} \right\}
\delta\left({\bf \tilde{P}}^2 - {\bf \tilde{M}}^2 \right) {\rm
sgn} \left({\bf
\tilde{P}}_0 \right).
\end{equation}
The $\delta$ Function in (\ref{eq31}) represents the defining equation
for the quasi--particle--energies
\begin{equation}\label{eq32}
E_{1/2} = Re \Sigma_0^r \pm \sqrt{\left(p_I - Re\Sigma_I^r
\right)^2 + \left(\kappa_o + Re \Sigma_I \right)^2}
\left.\right|_{\omega=E_{1/2}}
\end{equation}
This is a nonlinear equation by the many particle
influence and shows immediately the particle/antiparticle symmetry.
 Therefore the single particle and antiparticle states
become dressed by the medium and yield a natural
generalization of the Dirac--Bruckner Theory. It can be seen that
(\ref{eq32}) is determined only by the retarded self energy,
which yields a mass off-shell behavior of the quasiparticles.
This is important to note for the discussion in chapter 4.

Let us now consider the meson equations, which read for the
retarded Green's function written in operator notation, where integration
about inner variables is assumed
\begin{equation}\label{eq33}
\left(\Box + m_0^2 \right) \left(-d^r \right) = 1 - \, \,
\Pi^r d^r.
\end{equation}
Moreover we have for the correlation function
\begin{equation}\label{eq34}
\left(\Box + m^2_0 \right)
\left(-d\right)^{\grlo}
= -\Pi^r \, d^{\grlo} \,
-\,\Pi^{\grlo} \, d^a.
\end{equation}
Combining (\ref{eq33}) and (\ref{eq34}) we obtain the following
important relation
\begin{equation}\label{eq35}
-d^{\grlo} = d^r \, \Pi^{\grlo}
\, d^a
\end{equation}
which presents the generalized optical theorem
\protect\cite{D90,KB62}.
Following the same arguments as deriving (\ref{eq26}) one gets
for the product
$d^rd^a$ in gradient expansion
\begin{equation}\label{eq36}
d^rd^a = \frac{1}{\left(\omega^2-E^2_d \right)^2 +
\left(\frac{\Gamma_\pi}{2} + 2 \varepsilon \omega \right)^2}.
\end{equation}
Here the quasi particle energy of mesons is introduced by
\begin{equation}\label{eq37}
E^2_d = p^2 + m^2 - Re \Pi.
\end{equation}
The spectral function of mesons in the case of nearly vanishing damping reads
\begin{equation}\label{eq39}
B = i \left(d^r - d^a \right) \longrightarrow 2 \pi \delta
\left(\omega^2 - E^2_d \right) {\rm sgn} (\omega) \qquad for
\qquad \Gamma \rightarrow 0.
\end{equation}

In static case with vanishing
damping one obtaines from (\protect\ref{eq36}) the Fourier--transformed
Yukawa--potential for the meson exchange
\begin{eqnarray}\label{eq38}
d^rd^a&\approx& \frac{1}{g^4 (4\pi)^4} V^2(p) \nonumber \\
V(r) &=& g^2 \frac{e^{-r/r_0}}{r}   \qquad
r_0^{-2}=m^2 - Re \Pi.
\end{eqnarray}
When $\Pi$ is determined by (\ref{eq17}) we have the possibility to
construct a self consistent system including fluctuation phenomena.

\section{Kinetic equations}\label{ch}
The starting point are the generalized KB equations
\protect\cite{JW84} which are exact and read in the space--time
notation
\begin{equation}\label{eq40}
\left[Re G^{r^{-1}} , \,\, G^{\stackrel{>}\!\!\!\!{<}} \right] -
\left[\Sigma^{\stackrel{>}\!\!\!\!{<}} \,\, , \,\, Re \, G
\right] = \frac{1}{2} \left\{ G^< \, \Sigma^> \right\}
- \frac{1}{2} \left\{ G^> , \, \Sigma^< \right\}.
\end{equation}
Here we used operator notations, where the integration is assumed
over inner variables and the brackets sign the commutator $[]$ and
anticommutator $\{ \}$  respectively.

Using the gradient expansion here we have a different physical
content than the one yielding (\ref{eq26}) and (\ref{eq33}). If we
suppose here
\begin{equation}\label{eq41}
G(X,p) \gg \left| \frac{\partial}{\partial X^\mu}
\frac{\partial}{\partial p_\mu} G(X,p) \right| \gg...
\end{equation}
we have to demand that $\Delta X \Delta p \gg 1$. This would
be justified in the case of one particle systems only by
classical description. Because we have a many particle system G
contains information averaged over space--time cells, which can
be in principle smaller than the one determined by the single particle
de Broglie wave length \protect\cite{MK79}.
Because in nuclear physics time gradients occur, which are in
principle nonvanishing, this assumption is the most restrictive
one and is now under investigation and will soon be published.We
have from (\ref{eq40}) in $p\omega RT$ variables
\begin{equation}\label{eq42}
\left[ Re \, G^{r^{-1}}, \,\, G^< \right] = G^< \,\, \Sigma^> - G^>
\Sigma^<
\end{equation}
where the bracket means the Poisson-bracket:
\begin{equation}\label{eq43}
\left[A,B \right] = \frac{\partial}{\partial \omega} A
\frac{\partial}{\partial T}B - \frac{\partial}{\partial \omega} B
\frac{\partial}{\partial T}A + \bigtriangledown_R A
\bigtriangledown_p B - \bigtriangledown_p A \bigtriangledown_R B.
\end{equation}
As a first step of approximation we neglect higher vertex
corrections in agreement with the gradient expansion and derive
from (\ref{eq18}) for the vertex function
\begin{equation}\label{eq44}
\Gamma_{a\bar{a}d}\left(\stackrel{-}{1} \,\, \stackrel{=}{1} \,\,
2\right) \approx \delta_{a\bar{a}} (1-\bar{1})
\delta_{ad}(\bar{1}-2).
\end{equation}
In the framework of this relativistic Random Phase Approximation
(RRPA) we get from
(\ref{eq11}), (\ref{eq12}), (\ref{eq17}), (\ref{eq20}),
(\ref{eq35}) and (\ref{eq44}) the complete set of equations
\begin{eqnarray}\label{eq45}
\left[G^{r^{-1}} \,\, , G^< \right](p\omega RT) &=& G^< \Sigma^> - G^> \Sigma^>
  \nonumber\\
\Sigma^{\grlo} (xX) &=& - ig^2_0  \,
G^{\grlo} \left(1 \, \stackrel{=}{1} \right)
 \, d^{\grlo} \left(\stackrel{=}{1} \, 1
\right)   \nonumber\\
d^{\grlo}(p\omega RT) &=& \frac{1}{\left| 4 \pi g^2_0
\right| ^2} V^2 (p,\omega) \Pi^{\grlo}
 \nonumber\\
\Pi^{\grlo}(xX) &=& ig^2 \, Tr
\left\{G^{\grlo}
G^{\grlo} \right\}.
\end{eqnarray}

It has to be stressed that the dynamical potential
introduced by the optical theorem (\protect\ref{eq36}) now contains
an infinitesimal sum of nucleon fluctuations by the polarization
function resulting in an effective meson exchange mass. Therefore
this approximation can be considered as a modified first Born
approximation including collective effects, especially density
fluctuations.

The set of equations (\ref{eq45}a-d) forms a closed set and
determines the correlation function $G^{\grlo}$
and therefore the kinetic description proposed we find a
connection between $G^{\grlo}$.

Therefore we proceed to the kinetic level of description by introducing the
generalized distribution
\begin{eqnarray}\label{eq46}
G^{>}(p\omega RT) &=& -iA(p\omega RT)(\, 1 - f(p\omega RT)\,)
\nonumber\\
G^{<} \,\,(p\omega RT) &=& i\, A(p\omega RT)\, f(p\omega RT)
\end{eqnarray}
which is until now an exact variable change without
approximations. Especially it fulfills the conditions
(\ref{eq30}).

As far as the generalized distribution should describe physical
particles and antiparticles we use the Dirac interpretation in
the quasiparticle picture. There an empty state of particle with
negative energy equals to an antiparticle state with positive
energy and we can write
\begin{equation}\label{eq47}
f(-\omega) = 1 - {F} (\omega),
\end{equation}
where $ F $ signs the antiparticle distribution. In
equilibrium this is identical with the fact, that the chemical
potential has to be chosen with opposite signs for particles and
antiparticles, which follows immediately from the conserved baryon
density \protect\cite{SW86}.\\
With the help of (\ref{eq31}) and (\ref{eq32}) we obtain for $G^{<}$
\begin{eqnarray}\label{eq48}
G^{<} = \frac{i\pi}{\sqrt{{\bf{P}}_1^2 + {\bf{M}}^2 }}
\left\{ \right.
\left(\gamma^0 E_1 - \gamma^1 {\bf P}_1 + {\bf M} \right) \,\,
{\rm f}\left(E_1 \right) \delta \left(\omega - E_1 \right)
\nonumber\\
- \left( -\gamma^0 E_2 - \gamma^1 {\bf P}_1 + {\bf M}\right)
{\rm f} \left(-E_2 \right) \delta \left(\omega + E_2 \right)
\left. \right\}.
\end{eqnarray}
The choice of (\ref{eq47}) coincides with Bezzerides and DeBois
\protect\cite{BB72} if we set $F^{<}(\omega) = f(\omega)$
and $F^>(\bar{\omega}) = 1- F (\omega)$.

If we now introduce (\ref{eq47}) and (\ref{eq48}) into the
kinetic equation (\ref{eq45}a) we have to perform the Poisson
brackets on the left side carefully. After partial integration
one can show, that the renormalization denominator, which arise from
the spectral function (\ref{eq31}) cancels exactly with the
factors following from the gradient expansion on the left side so
that the drift term takes the form
\[
\frac{\partial}{\partial T}f + \bigtriangledown_R E
\bigtriangledown_p f - \bigtriangledown_p E \bigtriangledown_R f.
\]
 we want to
restrict to virtual meson exchange and neglect all density terms
in the meson Green's function. After somewhat extensive but
straight forward calculation using the set of equations
(\ref{eq45}a-d) we finally arrive at the kinetic equation for the
particle distribution, if we integrate both sides over
positive frequencies. The equation for antiparticles are obvious
by replacing ${\rm f}_p \longleftrightarrow F_p$.

We use for
shortness the following abbreviation of the
dynamically potential (\ref{eq36}) without static
approximation
\begin{eqnarray}\label{49}
\lefteqn{
V^2 (p,{\bar p},p_1,{\bar p1})=}\nonumber\\
&&\left (2 (p {\bar p_1})({\bar p}p_1)+ 2 (p p_1)({\bar p}{\bar p_1})-
2\kappa^2 [(p_1{\bar p_1})+(p{\bar p})]+4 \kappa^4 \right ) V_v^2
\left({\bar p}-p \right) \nonumber\\
&-&(p{\bar p} +\kappa^2)
(p_1{\bar p_1} +\kappa^2) V^2_s \left({\bar p}-p \right)
\end{eqnarray}
As already mentioned following (\ref{eq20})
no mixed terms between $g_r$, $g_s$ occur in the cross sections
due to the additive behaviour
of the different self energies.
The kinetic equation reads now
\begin{eqnarray}\label{kine}
\frac{\partial}{\partial T}f &+& \bigtriangledown_R E
\bigtriangledown_p f - \bigtriangledown_p E \bigtriangledown_R f
= Tr \int \left[g^< \Sigma ^> - g^> \Sigma^< \right]
\frac{d\omega}{2\pi} \, + \nonumber\\
&+& \frac{\partial}{\partial T} Tr \int P \frac{1}{(\omega -
\omega')^2} \left[ g^< \Sigma^> -g^> \Sigma^< \right]
\frac{d\omega}{2\pi} \frac{d\omega'}{2\pi}
\end{eqnarray}
with the collision integral up to first {\it two} order gradient expansions.
The second part of the right side of (\protect\ref{kine}) is obtained by
the second order gradient expansion or equivalently by the expansion of the
time retardation up to first order. It ensures the complete energy conservation
\protect\cite{GKL88},
whereas the first part of (\protect\ref{kine}) ensures only the quasiparticle
energy.
Because we restrict here to the case, where the quasiparticle picture is valid,
only the first part will be discussed. In a later paper we will present the
results obtained
for the second part.
The first part of the right side from (\protect\ref{kine}) containes 8
processes.
The first one, which describes elastic particle particle scattering reads
\begin{eqnarray}\label{ela}
\lefteqn{
\int \frac{d\overline{p}^3}{(2\pi)^3}
\frac{dp_1^3}{(2\pi)^3}
\frac{d\bar{p}_1^3}{(2\pi)^3} \,(2\pi)^4 \delta^{(4)}
\left(\bar{p}_1 + \bar{p} - p_1 - p \right)
\frac{1}
{\left| E_1 \bar{E}_1 E \bar{E} \right|}*
}  \nonumber\\
&&V^{2}(p,{\bar p},p_1,{\bar p_1})
  \left\{ f_p f_{p_1} N\left(f_{\bar{p}} f_{\bar{p}_1}\right)
-f_{\bar{p}} f_{\bar{p}_1} N\left(f_p f_{p_1}\right) \right\}.
\end{eqnarray}
Here and in the following we used the Pauli blocking factors
\begin{eqnarray*}
N(F)&=& 1 - F \\
N(F \, f) &=& (1 - F) (1 - f)  \\
N(F \, f \, G) &=& (1 - F) (1 - f) (1 - G).
\end{eqnarray*}
The next two processes are the the crossing symmetric processes in the t and u
channel.
They are obvious by the crossing symmetric relations and describe the elastic
particle-
antiparticle scattering. The next 5 processes are only possible due to the
off-shell
character of the quasi-particle energies (\protect\ref{eq32})
\begin{eqnarray}\label{int2}
\lefteqn{
\int \frac{d\overline{p}^3}{(2\pi)^3}
\frac{dp_1^3}{(2\pi)^3}
\frac{d\bar{p}_1^3}{(2\pi)^3} \,(2\pi)^4
\frac{1}
{\left| E_1 \bar{E}_1 E \bar{E} \right|}*
}  \nonumber\\
\left [ \right .
&-&V^{2}(p,{\bar p},-p1,{\bar p1})
\delta^{(4)} \left({\bar{p}_1} + p_1 + {\bar{p}} - p \right)
\left\{ f_pN \left(f_{\bar{p}} F_{p_1} f_{\bar{p}_1} \right)
- f_{\bar{p}} f_{\bar{p}_1} F_{p_1}  \right\}
\nonumber\\
&-&V^{2}(p,{\bar p},p_1,-{\bar p_1})
\delta^{(4}) \left(-{\bar{p}_1} -p_1 + {\bar{p}} -p \right)
\left\{
f_p f_{p_1} F_{\bar{p}_1} - N \left(f_p f_{p_1} F_{\bar{p}_1} \right)
f_{\bar{p}}
\right\}
\nonumber \\
&-&V^{2}(p,{\bar p_1},p_1,-{\bar p})
\delta^{(4}) \left(-{\bar p} -p_1 + {\bar p_1} -p \right)
\left\{
f_p f_{p_1} F_{\bar{p}} - N \left(f_p f_{{\bar p}_1} F_{\bar{p}} \right)
f_{\bar{p}}
\right\}
\nonumber \\
&-&V^{2}(p,-{\bar p},-p_1,-{\bar p_1})
\delta^{(4}) \left(-{\bar{p}}_1 + p_1 - {\bar{p}} - p \right)
\left\{ f_p F_{\bar{p}} F_{\bar{p}_1} - F_{P_1} \, N \left(fp
F_{\bar{p}} F_{\bar{p}_1} \right)  \right\}
\nonumber\\
&+&V^{2}(p,-{\bar p},p_1,-{\bar p_1})
\delta^{(4}) \left(-{\bar{p}}_1 - p_1 - {\bar{p}} - p \right)*
\nonumber\\
&& \left . \qquad \qquad \left\{
f_p F_{\bar{p}} f_{p_1} F_{\bar{p}_1} - \left(1 - f_p \right)
\left(1 - F_{\bar{p}} \right) \left(1 - f_{p_1} \right)
\left(1 - F_{\bar{p}_1} \right)  \right\} \right ] . \nonumber \\
\end{eqnarray}
Analyzing the energy momentum conserving $\delta$-function on finds in
connection with
(\protect\ref{eq32}) that these processes occur if
the c.m. energy
$\sqrt{s}$ fulfills the condition
\begin{equation}\label{condi}
\sqrt{s}=-2 Re \Sigma^r_o.
\end{equation}
This means, that in a nucleus system, where the zero part of the selfenergy
become negative, new reaction channels will be opened by many particle
influence.
One may think on pseudoscalar and pseudovector meson coupling, where the
absence
of mean field terms of the vector mesons ensures the condition
(\protect\ref{condi}).

Further, it can be seen explicitly that the relativistic treatment yields
some forefactors (\protect\ref{49}) to
the dynamical potentials
(\ref{eq36}). The dynamical
potentials have to be determined selfconsistently with the
kinetic equation by means of (\protect\ref{kine}) and the
polarization function (\protect\ref{eq45}d).

Therefore we arrived
at a generalized relativistic Lenard-Balescu equation. The dynamical potentials
reflect the influence of virtual mesons and are written for each
process with the influence of the dynamical density fluctuations by
(\ref{eq36}).

The first
3 terms are presenting the elastic scattering between particles
and antiparticles respectively. The next 5 processes are only
possible, if the particles are off shell, which is ensured by the
quasiparticle energies shown in (\ref{eq32}). This can
be understood as particle creation and destruction by means of
virtual meson exchange. It is important to recognize, that these
processes are special effects of many particle treatment and
are forbidden by momentum and energy conserving
$\delta$-functions in the case of vanishing many particle
influence.

The equation (\ref{kine}) possesses all properties of a
selfconsistent quantum--mechanical kinetic equation which describes the
relativistic behaviour of a meson-nucleon system from the many
particle point of view. Particularly, it contains Pauli blocking
and inelastic processes by the medium.
This
is the main result of the present paper and should be
the starting point to kinetic description of hot nuclear matter.

To complete we give the in medium cross sections obtained for the elastic
process
(\protect\ref{ela}). For shortness we denote $Re \Sigma_o^r=\Delta$ and
$\tilde{\kappa}=\kappa+Re \Sigma_i^r$. The proton-proton cross section reads
\begin{eqnarray}\label{crosss}
\lefteqn{
\frac{d \sigma}{d \Omega}=\frac{1}{32 \pi^2} \left \{ \right.
}\nonumber\\
&&\frac{2 g_v^4}{s (m_v^2-t)^2} \left[ (\Delta (\sqrt{s}+\Delta)+
\tilde{\kappa}^2
-\frac{u}{2})^2+(\Delta (\sqrt{s}+\Delta)-\tilde{\kappa}^2
+\frac{s}{2})^2 \right . \nonumber \\
&\qquad& \left . -2 \tilde{\kappa}^2 (\Delta (\sqrt{s}+\Delta)-\frac{t}{2})
\right] \nonumber\\
&-&\left .
\frac{2 g_s^4}{s (m_s^2-t)^2} (\Delta (\sqrt{s}+\Delta)+2 \tilde{\kappa}^2
-\frac{t}{2})^2 \right \}.
\end{eqnarray}
Here $s,t,u$ are the Mandelstam variables. This means that the in
medium cross sections are effected by the dressed Baryon mass and
an additional term proportional to the vector part of the self
energy.

The in-medium proton-antiproton elastic
cross sections as well as the discussed medium caused inelastic
ones can be derived from the corresponding collision integrals
(\protect\ref{ela}) and (\protect\ref{int2}) in the same way. It
turns out that they are identical to the cross sections one
obtaines from (\protect\ref{crosss}) by using crossing symmetric
relations in the t or u channel or to inelastic crossing. This
means that we end up with an expression for the medium dependent
cross section for nucleons, which possess crossing symmetries and
includs an
infinitesimal series of meson interactions by the RPA polarization function
resulting in an effective meson mass (\protect\ref{eq37}),

\section{Conclusion}\label{ch6}

We have analysed the structure of equations for a nonequilibrium
situation consisting in a relativistic nucleon-meson system. New
features arise due to the consequent nonequilibrium treatment as
pair creation and destruction and dynamically interacting meson masses.
With one
formalism of real time Green function one can determine the self
energy, the quasiparticle behaviour and kinetic equations of the
system and therefore the scattering measures. The kinetic
equations are derived of Lenard Balescu type by $V_s$
approximation which coincide with the RPA screened potentials in
nonrelativistic treatment. These obtained potentials may serve as
a starting point for constructing optical potentials
\protect\cite{MD90}. On the other hand in-medium cross sections can be
derived from the kinetic equation, which show no mixed terms of
mesons contributions. This follows
immediately from the more general decoupling of equations by the self energies.
In a forthcoming work we
will present the numerical results for the in-medium cross
sections derived from (\protect\ref{kine}) for the now opened inelastic
scattering channel.

\section{Acknowledgement}
One of the authors, K.M., like to thank Prof. Malfliet in Groningen for
hospitality, Prof. Muenchow, Dubna for stimulating discussions
and P. Henning in Darmstadt for useful hints.
\bibliography{kmsr3.bib,kmsr2.bib,kmsr1.bib,kmsr.bib}
\bibliographystyle{unsrt}

\end{document}